\pdfoutput=1
\documentclass[acmtog,nonacm]{acmart}
\setcopyright{none}
\usepackage{array,amsmath,booktabs,graphicx,algorithm,algpseudocode,xcolor,multirow}
\newcommand{\ManuscriptAssets}{assets}

\begin{document}
\title{Direction-Guided Integer Cone Optimization via Hodge Decomposition}
\author{Zhantao Liu}
\email{zhl162@ucsd.edu}

\begin{abstract}
Turning a dense cross field $G$ into a surface parameterization requires choosing discrete cones and holonomy around non-contractible loops before integration. Cross-edge variation in $G$ mixes smooth fanning, which conformal scaling absorbs without rotating the grid, and pure swirl, which forces grid deviation. Scale-only cone selectors ignore $G$ altogether, and guided smoothers penalize both effects equally. We score candidate integer matchings with a discrete conformal response model based on a Hodge projection. Its geometry-only operators factor once per mesh and orthogonally separate conformal log-scale $u$, angle correction $\gamma$, and harmonic handle residual $h$. Reusing these systems scores multi-edge and nonlocal cone edits globally without nested nonlinear solves. The same decoupled model also supports regional cone exclusion, local directional edits, and sizing targets. Across 278 QuadWild meshes evaluated end-to-end with Rectangular Surface Parameterization (\textsc{RSP}), our default setting yields 211 valid maps versus 195 for direct rounding of $G$. On the 192 jointly valid meshes, paired medians trade $1.41^\circ$ higher angular error for two fewer cones and $0.040$ lower log-scale distortion. On closed genus-$0$ meshes within the cone budgets of scale-only selectors at their default distortion target, our scale-priority setting lowers angular error on 64 of 77 meshes (median reduction $1.43^\circ$) with near-zero median log-scale change.
\end{abstract}

\ccsdesc[500]{Computing methodologies~Mesh geometry models}
\keywords{surface parameterization, cross fields, cone singularities, discrete conformal geometry, integer optimization}

\begin{teaserfigure}
  \centering
  \includegraphics[width=\textwidth]{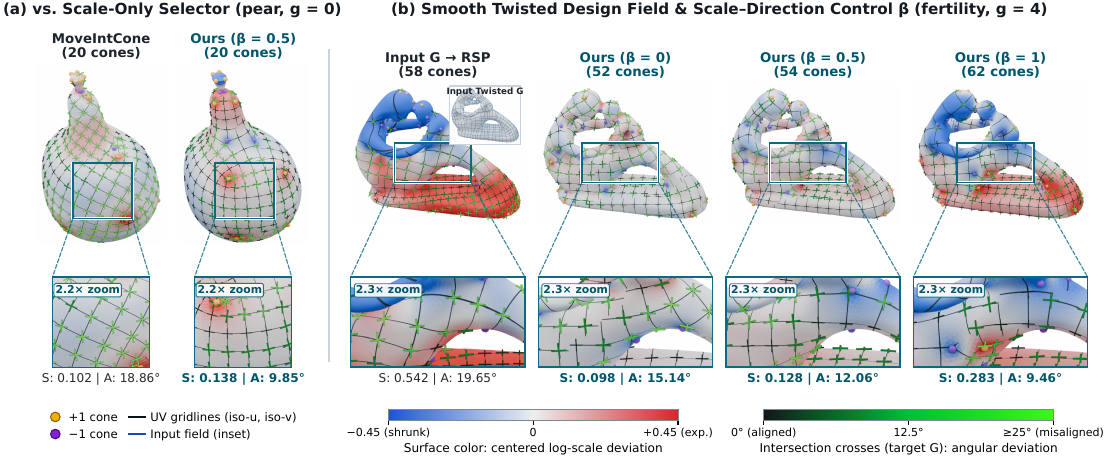}
  \caption{All grids use the same parameterization solver~\cite{Corman2025RSP}. \textbf{(a)}~With $20$ cones in each map, direction-guided cone placement nearly halves the angular error at higher scale distortion. \textbf{(b)}~All three optimized maps improve both alignment and scale over the leftmost input topology, and increasing the alignment preference $\beta$ trades lower angular error for greater scale distortion.}
  \Description{Six 3D surface parameterizations in two groups. Left: two views of the pear mesh with 20 cones each, comparing MoveIntCone against our method with magnified insets showing closer cross alignment for our method. Right: four views of the genus-4 fertility mesh under a height-twisted field, comparing direct Input G to RSP against our method at beta equals 0, 0.5, and 1 with magnified insets showing progressively tighter cross alignment as beta increases.}
  \label{fig:teaser}
\end{teaserfigure}

\maketitle

\section{Introduction}
\label{sec:intro}

A target cross field $G$ prescribes preferred grid directions on each triangle, but leaves a discrete choice across every shared edge: which of a cross's four equivalent arms continues onto the neighboring face. Accumulating these quarter-turn matchings around vertices determines the cone singularities, while summing them around handles sets the holonomy around non-contractible loops. Together, they fix the topological class available to a parameterization on closed surfaces (Fig.~\ref{fig:teaser}a--b) and open shells. In Fig.~\ref{fig:teaser}a, both $20$-cone grids are computed by the same Rectangular Surface Parameterization (\textsc{RSP}) solver~\cite{Corman2025RSP} against the same target $G$. Scale-only cones from Move\-Int\-Cone~\cite{Du2026EfficientIntegerCones} deviate by $18.86^\circ$ with log-scale distortion $0.102$, whereas our balanced default lowers angular error to $9.85^\circ$ at a higher log-scale distortion of $0.138$. What geometric signal in $G$ distinguishes two cone configurations of the same size?

Consider how cross directions vary from face to face (Fig.~\ref{fig:decomposition}a--b). Given suitably placed cones, conformal rescaling of cell sizes can match smooth fanning or bending of $G$ with zero grid rotation. Pure swirl cannot be absorbed by scaling and instead forces the grid to rotate away from $G$ unless relieved by singularities. Existing selectors miss this separation in different ways: scale-only methods~\cite{Soliman2018OptimalCones,Zhang2024PracticalIntegerCones,Du2026EfficientIntegerCones} ignore $G$ and place identical cones for every directional design. Scale-agnostic guided relaxations~\cite{Knoppel2013GloballyOptimal} take $G$ into account, but penalize all cross-edge turning uniformly and conflate conformal fanning with swirl. \citet{Myles2013ControlledDistortion} relate cross-field variation to conformal scale, but inherit the input cones wherever directions are prescribed. \textsc{RSP}~\cite{Corman2025RSP} instead discovers singularities through progressive cut-and-glue interleaved with full-mesh nonlinear solves (Section~\ref{sec:related}).

We evaluate candidate integer edge matchings using a fixed \emph{discrete conformal response model} (Section~\ref{sec:model}) that predicts scale variation and face rotations relative to $G$. A discrete Hodge--Helmholtz projection orthogonally decomposes cross-edge discrepancy into conformal log-scale $u$ for smooth fanning, angle correction $\gamma$ for unabsorbable swirl, and a harmonic residual $h$ for handle mismatch (Fig.~\ref{fig:decomposition}c). Because these operators depend only on mesh geometry, we factor them once per mesh and use a single weight $\beta$ to balance scale uniformity against directional alignment. A lightweight smooth fit selects consistent per-face $90^\circ$ orientations and exports a smooth starting frame for parameterization.

Moving one cone changes scale and alignment across the whole mesh, making nested nonlinear parameterizations impractical during search. Because all continuous responses satisfy fixed linear systems, we reuse the prefactored operators to score multi-edge and nonlocal edits directly (Section~\ref{sec:opt}). The same factorizations also support regional cone exclusion, local directional edits, and relative sizing targets (Fig.~\ref{fig:controls}). We evaluate every selected topology end-to-end through \textsc{RSP} on 278 QuadWild meshes~\cite{Pietroni2021Reliable}, spanning up to 15 boundary loops and genus 133, plus 50 raw Thingi10K meshes~\cite{Zhou2016Thingi10K} (Section~\ref{sec:eval}). Across these benchmarks, our cones improve validity and scale uniformity over direct rounding and scale-agnostic smoothing, and reduce angular error at matched cone budgets relative to scale-only selectors.

\section{Related Work}
\label{sec:related}

\paragraph{Joint Connection and Integrability Methods.}
Methods coupling connections with distortion or integrability come closest to ours, yet treat dense guidance differently. Trivial connections~\cite{Crane2010TrivialConnections} build flat connections for prescribed singularities without choosing them. \citet{Myles2013ControlledDistortion} combine a connection-form Hodge decomposition with log-scale distortion and smoothness for curve-aligned progressive rounding and cone relocation, but inherit the input field's cones wherever directions are prescribed throughout. Adding the orthogonal angle-correction potential $\gamma$ lets us jointly trade off conformal scale and dense directional fidelity to re-optimize cones and loop holonomies everywhere. \citet{Ma2025IterativeSingularities} pair local moves and global matching under sparse hard constraints. \citet{Couplet2026SurfaceOdeco} optimize continuous orthogonally decomposable (odeco) tensor fields with optional alignment and sizing terms, rather than searching discrete integer cones and holonomies via linear move costs. \textsc{RSP}~\cite{Corman2025RSP} discovers cones via progressive cut-and-glue with repeated nonlinear solves (Section~\ref{sec:eval-baselines}); we instead decouple topology selection into three reusable linear systems. Non-orthogonal frame and metric methods~\cite{Panozzo2014FrameFields,Diamanti2015IntegrablePolyVector,Jiang2015MetricCustomization,SagemanFurnas2019ChebyshevNets} absorb curl by relaxing orthogonality or equal scale, whereas we target orthogonal cross fields and rectangular maps.

\paragraph{Curvature-Driven Conformal Cones.}
Prescribed cones reduce area distortion in circle-pattern maps~\cite{Kharevych2006CirclePatterns}. Vertex scaling~\cite{Springborn2008ConformalEquivalence,BenChen2008ConformalFlattening} and incremental flattening~\cite{Myles2012IncrementalFlattening} also link cone placement to metric distortion. \citet{Soliman2018OptimalCones} formulate cone selection as a sparse inverse-Poisson problem under an $L^2$ Hencky norm. Subsequent methods construct sparse cones under distortion bounds~\cite{Fang2021SparseCones,Li2023EfficientCones}, impose integer cone angles~\cite{Li2022SparseIntegerCones,Zhang2024PracticalIntegerCones}, move integer cones~\cite{Du2026EfficientIntegerCones}, or use meromorphic quartic differentials~\cite{Wang2026ClosedFormCones}. A flattener such as \textsc{BFF}~\cite{Sawhney2017BFF} can realize prescribed cones. Rotating $G$ leaves a scale-only score unchanged, even when it changes how well the resulting map follows $G$.

\paragraph{Frame Fields, Holonomy, and Seamless Maps.}
Directional field design~\cite{Vaxman2016DirectionalSurvey} supplies guidance for quad meshing~\cite{Bommes2013QuadSurvey}. Early methods edit field singularities~\cite{Palacios2007RotationalFields} or suppress those induced by fine geometric detail~\cite{Ray2009GeometryAware}. Smooth cross-field solvers~\cite{Knoppel2013GloballyOptimal} support optional guidance, whereas \textsc{IOQ}~\cite{Farchi2018IntegerOnly} eliminates continuous angles to optimize closed-surface integer holonomy via resistance-distance costs without an external target $G$. Scale-agnostic guided methods using quadratic smoothing~\cite{Knoppel2013GloballyOptimal}, circle-bundle minimal sections~\cite{Palmer2024MinimalSections}, semidefinite synchronization~\cite{PachecoTallaj2026Synchronizing}, or learned generators~\cite{Dong2025CrossGen,Zhang2026LearningSingularities} optimize cross-field smoothness without accounting for conformal scale variation. Our projection instead separates the turning that conformal scaling absorbs from the rotation that remains. Downstream, harmonic forms on a singularity graph generate quad layouts~\cite{Tong2006HarmonicQuads}, field-aligned parameterizers build seamless or integer-grid maps~\cite{Kalberer2007QuadCover,Bommes2009MIQ,Bommes2013IGM,Myles2014RobustFieldAligned}, and holonomy solvers realize prescribed cones or signatures~\cite{Campen2020ArbitraryCones,Shen2022HolonomySignatures}; making seam translations integral is a separate quantization step~\cite{Campen2015QuantizedGlobal}. Our search chooses direction-aware cones and holonomies upstream.

\section{Discrete Conformal Response Model}\label{sec:model}

\begin{figure*}[t]
  \centering
  \includegraphics[width=0.90\textwidth]{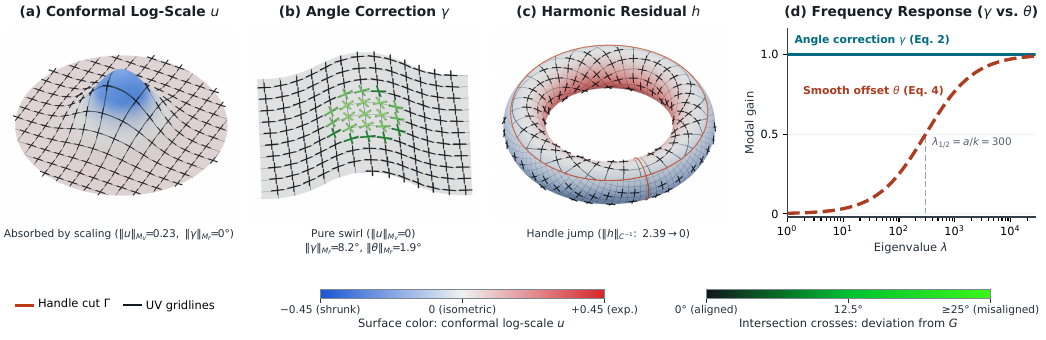}
  \caption{Hodge decomposition of the cross-edge discrepancy in Eqs.~\eqref{eq:hodge}--\eqref{eq:ortho} and frequency response in Eq.~\eqref{eq:smooth}. \textbf{(a)}~Smooth bending or fanning is absorbed with zero rotation by conformal log-scale $u$. \textbf{(b)}~Pure swirl registers solely as angle correction $\gamma$, whereas the smooth offset $\theta$ is strongly attenuated. \textbf{(c)}~On a genus-$1$ torus (shown pre-shift), mismatched loop holonomy around a handle leaves a global harmonic residual $h$ and a grid mismatch across $\Gamma$, both eliminated by a holonomy shift $\Delta p$. \textbf{(d)}~On pure-swirl eigenmodes, the angle correction has unit gain, while the smooth offset attenuates low frequencies.}\label{fig:decomposition}
  \Description{Four-panel diagram of the Hodge decomposition and frequency response: (a) conformal log-scale u absorbing smooth field bending without rotation, (b) pure swirl with zero conformal log-scale, 8.2-degree RMS angle correction gamma, and 1.9-degree RMS smooth offset theta, (c) harmonic residual h on a genus-1 torus eliminated by a loop holonomy shift, and (d) modal gain curves versus eigenvalue lambda comparing constant unit gain for angle correction gamma against low-frequency attenuation with half-gain cutoff lambda equals 300 for smooth offset theta.}
\end{figure*}

We score candidate cones and non-contractible loop holonomies via a linear conformal response model on closed meshes (Sections~\ref{sec:model-connection}--\ref{sec:model-smooth}), then extend it to open boundaries (Section~\ref{sec:model-boundary}) and spatial design controls (Section~\ref{sec:model-controls}); Table~\ref{tab:notation} collects our notation.

\begin{table}[t]
\centering
\caption{Core notation for Sections~\ref{sec:model}--\ref{sec:opt}.}\label{tab:notation}
\footnotesize
\setlength{\tabcolsep}{2.5pt}
\renewcommand{\arraystretch}{1.0}
\begin{tabular}{@{}l >{\raggedright\arraybackslash}p{0.63\columnwidth}@{}}
\toprule
Symbol & Meaning \\
\midrule
$V, E, F;\, V_I, E_I$ & Vertices, edges, and faces; interior subsets ($V_I{=}V,\, E_I{=}E$ if $b{=}0$) \\
$\alpha,\, p,\, z_I$ & Target face angles ($\bmod\,q$, $q{=}\pi/2$), integer matching, unit interior cone charges \\
$u,\, \gamma,\, \theta$ & Conformal log-scale, angle correction, smooth offset \\
$r(p,\alpha),\, h,\, \omega$ & Cross-edge discrepancy, $2g$-D harmonic residual, smoothed cross-edge residual \\
$B,\, B_I,\, D$ & Vertex--edge incidence (interior $B_I$), dual face-difference ($B_I D{=}0$) \\
$R,\, A$ & Boundary-loop scale map ($u{=}Rs$; $R{=}I$ if $b{=}0$), reduced gradient ($A{=}B^\top R$, $A^\top D{=}0$) \\
$C,\, W,\, M_V,\, M_F$ & Clamped cotangent Hodge star, positive barycentric dual weights, unit-area masses \\
\bottomrule
\end{tabular}
\end{table}

\subsection{Discrete Connections and Admissible Singularities}\label{sec:model-connection}

Let $\mathcal{M} = (V,E,F)$ be an oriented closed triangle mesh of genus $g$ ($\chi = 2 - 2g$), with signed vertex--edge incidence matrix $B \in \{-1,0,1\}^{|V|\times|E|}$ and dual face-difference operator $D \in \{-1,0,1\}^{|E|\times|F|}$ oriented consistently so that face jumps telescope around every vertex ($BD = 0$). In local orthonormal face frames, a dense unit target cross field $G$ is represented by face angles $\alpha \pmod{q}$ ($q = \pi/2$). Across each oriented dual edge, an integer matching $p \in \mathbb{Z}^{|E|}$ selects which arm of the neighboring cross continues across the edge~\cite{Ray2008NSymmetry}. Together with Levi-Civita parallel transport $t$ and vertex angle-defect curvature $K$, $p$ determines the cross-edge discrepancy $r(p,\alpha)$ and integer cone charges $z$:
\begin{equation}\label{eq:connection}
  r(p,\alpha) = D\alpha - t - qp, \qquad z = \frac{K + Bt}{q} + Bp.
\end{equation}
Local frame holonomy makes $(K + Bt)/q$ integer-valued, so $z$ is integer for any integer $p$. We restrict singularities to unit cones $z_v \in \{-1,0,1\}$, corresponding to $\{5,4,3\}$-valent quad vertices (cone angle $2\pi - qz_v$, cone count $\|z\|_1$, $\sum_v z_v = 4\chi$). For genus $g \ge 1$, matchings with identical $z$ can still differ in their $2g$ \emph{generator holonomies} ($B\Delta p = 0$): the signed sums of quarter-turns around a basis of non-contractible dual loops, counting full $2\pi$ turns rather than reducing modulo $2\pi$. Together, cones and generator holonomies lock the topological class prior to UV integration.

\subsection{Hodge Decomposition of the Cross-Edge Discrepancy}\label{sec:model-response}

Away from cones, conformally rescaling the metric to $e^{2u}\mathsf{g}$ shifts the Levi-Civita connection by the rotated differential $\star du$~\cite{Crane2010TrivialConnections,Myles2013ControlledDistortion,Crane2013DEC}. On the mesh, we form the positive diagonal cotangent Hodge star $C \succ 0$~\cite{Desbrun2005DiscreteForms} from the opposite angles $\varphi_e, \psi_e$. Conformal scaling then shifts transport to $\tilde{t} = t - CB^\top u$, while deflecting face angles to $\tilde{\alpha} = \alpha + \gamma$ leaves residual $\tilde{r} = D\tilde{\alpha} - \tilde{t} - qp = r(p,\alpha) + D\gamma + CB^\top u$. Minimizing $\tfrac{1}{2}\|\tilde{r}\|_{C^{-1}}^2$ yields a discrete Hodge--Helmholtz projection~\cite{Polthier2003Hodge,Fisher2007TangentFields}. Because $BD = 0$ on closed meshes, $CB^\top u$ and $D\gamma$ are $C^{-1}$-orthogonal, so the normal equations decouple into two geometry-only sparse Poisson solves with minimum-norm harmonic residual $h = \tilde{r}_{\min}$:
\begin{equation}\label{eq:hodge}
\begin{gathered}
  r(p,\alpha) = -CB^\top u - D\gamma + h, \\
  (BCB^\top)u = -Br = qz - K, \qquad (D^\top C^{-1}D)\gamma = -D^\top C^{-1}r.
\end{gathered}
\end{equation}
With diagonal vertex and face mass matrices $M_V, M_F$ normalized to unit total area ($\mathbf{1}^\top M_V\mathbf{1} = \mathbf{1}^\top M_F\mathbf{1} = 1$), zero-mean gauges ($\mathbf{1}^\top M_V u = 0$ and $\mathbf{1}^\top M_F \gamma = 0$) remove constant nullspaces, so $\|u\|_{M_V}^2$ equals the spatial log-scale variance. On non-Delaunay edges, we clamp weights to $C_{ee} = \max\bigl(\tfrac{1}{2}(\cot\varphi_e{+}\cot\psi_e), 10^{-5}\bigr) > 0$; any positive diagonal $C \succ 0$ preserves $BD = 0$ orthogonality, and Section~\ref{sec:model-smooth} addresses local spikes on sliver triangles.

Eq.~\eqref{eq:hodge} splits $r(p,\alpha)$ into three $C^{-1}$-orthogonal terms:
\begin{equation}\label{eq:ortho}
\begin{gathered}
  \|r\|_{C^{-1}}^2 = \|CB^\top u\|_{C^{-1}}^2 + \|D\gamma\|_{C^{-1}}^2 + \|h\|_{C^{-1}}^2, \\
  Bh = 0, \qquad D^\top C^{-1}h = 0.
\end{gathered}
\end{equation}
Driven by $qz - K$, conformal log-scale $u$ absorbs smooth bending and fanning with zero rotation (Fig.~\ref{fig:decomposition}a). On a fixed angle branch, angle correction $\gamma$ gives the face rotations from $G$ to a minimum-variation field for matching $p$, capturing pure swirl that isotropic scaling cannot absorb (Fig.~\ref{fig:decomposition}b). Perturbing $\alpha \mapsto \alpha + \delta$ at fixed $p$ leaves $u$ and $h$ unchanged while shifting $\gamma \mapsto \gamma - \Pi_F\delta$, where $\Pi_F = I - \mathbf{1}\mathbf{1}^\top M_F$ removes the area-weighted mean. Changing the target can therefore change which matching has the lower directional score. Harmonic residual $h \in \mathcal{H}$ records the $2g$-dimensional handle mismatch unremovable by single-valued $(u,\gamma)$ (Fig.~\ref{fig:decomposition}c; $h \equiv 0$ for $g = 0$). Rotating selected face reference arms via a cone- and holonomy-preserving \emph{per-face branch shift} $p \mapsto p - Dn$ ($n \in \mathbb{Z}^{|F|}$) leaves $u$ and $h$ invariant yet shifts $\gamma \mapsto \gamma - q\Pi_F n$, so scoring $\|\gamma\|_{M_F}^2$ first requires aligning $p$ to a consistent $90^\circ$ angle branch.

Rather than the edge Dirichlet energies in Eq.~\eqref{eq:ortho}, we penalize mass-weighted potential norms $\|u\|_{M_V}^2$ and $\|\gamma\|_{M_F}^2$. Because $u$ has a logarithmic singularity at each cone, $\|CB^\top u\|_{C^{-1}}^2$ depends on mesh resolution, whereas the $L^2$ Hencky norm $\|u\|_{M_V}^2$ stays finite and resolution-consistent~\cite{Soliman2018OptimalCones}; $\|\gamma\|_{M_F}^2$ directly measures area-weighted deviation, and varying preference weights never alters any candidate's $(u,\gamma,h)$ split.

\subsection{Branch Alignment, Scoring, and Warm-Start Export}\label{sec:model-smooth}

Two practical needs remain beyond Eq.~\eqref{eq:hodge}: selecting a consistent per-face angle branch before evaluating $\|\gamma\|_{M_F}^2$, and exporting a smooth warm-start frame free of local spikes that $C_{ee} \ge 10^{-5}$ can induce in $\gamma$ on narrow slivers. Both are handled by an auxiliary smooth fit $(\theta,\omega)$ under positive barycentric dual weights $W \succ 0$:
\begin{equation}\label{eq:smooth}
\begin{gathered}
  F_p(\theta) = \frac{a}{2}\|\theta\|_{M_F}^2 + \frac{k}{2}\|r + D\theta\|_W^2, \\
  A_\theta \theta = -k D^\top W r, \qquad \omega = r + D\theta,
\end{gathered}
\end{equation}
where $A_\theta = k D^\top W D + a M_F \succ 0$, $\omega$ is the smoothed cross-edge residual of $\alpha + \theta$, and $W_{ee} > 0$ divides primal edge length by the sum of distances from its midpoint to the two adjacent face centroids.

Alternating Eq.~\eqref{eq:smooth} with per-face branch shifts $p \leftarrow p - Dn$ ($n = \operatorname{round}(\theta/q)$) performs block coordinate descent on $F_p(\theta)$ jointly over continuous $\theta$ and integer shifts $n$, descending at each nonzero step and terminating finitely at a \emph{branch-aligned representative}. On this representative, we score
\begin{equation}\label{eq:score}
\begin{aligned}
  E(p) = &\;\frac{w_A}{2}\|u\|_{M_V}^2 + \frac{w_D}{2}\|\gamma\|_{M_F}^2 + \frac{w_H}{2}\|h\|_{C^{-1}}^2 \\
         &+ \frac{\lambda_s}{2}\|\omega\|_W^2 + w_C\|z\|_1,
\end{aligned}
\end{equation}
with defaults $a = 0.3$, $k = \lambda_s = 10^{-3}$, $w_H = 0.1$, $w_C = 7.5\times 10^{-4}$, and symmetric trade-off weights $w_A = 10^{0.5-2\beta}$, $w_D = 10^{2\beta-1.5}$, sweeping $w_A/w_D$ from $100{:}1$ ($\beta{=}0$) through $1{:}1$ ($\beta{=}0.5$) to $1{:}100$ ($\beta{=}1$); under unit-area normalization, $u$ and $\gamma$ are dimensionless $O(1)$ quantities. Search exports $(p, z_I, u, \alpha{+}\theta)$ ($z_I{=}z$ if $b{=}0$); \textsc{RSP}~\cite{Corman2025RSP} warm-starts from $(p, \alpha{+}\theta)$ with zero scale and optimizes against $G$.

\paragraph{Why score $\gamma$ rather than $(\theta,\omega)$?}
Omitting $u$ and minimizing $\|\omega\|_W^2$ penalizes smooth bending that conformal scaling absorbs. Even alongside $u$, scoring $\|\theta\|_{M_F}^2$ distorts directional sensitivity across scales (Fig.~\ref{fig:decomposition}d). On a nonconstant pure-swirl eigenmode $r = D\phi$ ($D^\top W D \phi = \lambda M_F \phi$), Eq.~\eqref{eq:hodge} recovers $\gamma = -\phi$ with unit gain across all frequencies, while Eq.~\eqref{eq:smooth} attenuates it to $\theta = -\tfrac{k\lambda}{a + k\lambda}\phi$. With dimensionless $W_{ee}$ and unit-area $M_F$, the cutoff $\lambda_{1/2} = a/k = 300$ is independent of global mesh scale. Still, $\theta$ damps low-frequency swirl into $\omega$ differently across meshes, so one weight on $\|\theta\|_{M_F}^2$ cannot calibrate consistently. Moreover, Eq.~\eqref{eq:hodge} decouples scale via $D^\top C^{-1}(CB^\top u) = (BD)^\top u \equiv 0$, whereas using $W \neq C^{-1}$ leaks scale gradients into $\theta$. We therefore score alignment with $\gamma$ so the physical terms rank candidates under one default weight. Alongside them, $\tfrac{\lambda_s}{2}\|\omega\|_W^2$ acts as a smoothness regularizer that trades ranking fidelity for search stability and flip-free parameterization (Section~\ref{sec:eval-controls}), and tiny sliver areas keep $\|\gamma\|_{M_F}^2$ insensitive to isolated spikes.

\subsection{Open Boundaries}\label{sec:model-boundary}

On open meshes with $b \ge 1$ boundary loops, $(p, r, D)$ live on interior edges $E_I$, and admissible cones and penalties restrict to interior vertices $V_I$ ($z_I \in \{-1,0,1\}^{|V_I|}$, $w_C\|z_I\|_1$). Boundary-vertex rows of $B$ no longer annihilate $D$ individually because incident interior edges form open fans. Setting $u = Rs$ assigns one log-scale to each interior vertex and one shared constant log-scale to each boundary loop; the reduced gradient $A = B^\top R$ contracts each loop into a virtual vertex so boundary fans telescope around every closed loop, restoring orthogonality $A^\top D = 0$ without prescribing unknown boundary geodesic curvatures:
\begin{equation}\label{eq:open}
\begin{gathered}
  A^\top D = 0, \quad (A^\top C A)s = -A^\top r, \quad u = Rs, \\
  (D^\top C^{-1}D)\gamma = -D^\top C^{-1}r, \quad h = r + D\gamma + CAs.
\end{gathered}
\end{equation}
This per-loop constant-scale condition preserves geometry-only factorization and matches the isotropic conformal boundary equations of \textsc{RSP} when boundary directions remain free, while downstream \textsc{RSP} can still vary boundary scale once anisotropic stretch activates. When conformal scale varies strongly along a boundary loop, however, one scale per loop can bias boundary distortion and add interior cones. Per-loop scales also separate the $2g$-dimensional harmonic residual $\mathcal{H}$ from the $(2g + b - 1)$-dimensional non-contractible loop holonomies $\ker B_I / \operatorname{im} D$; on an annulus ($g = 0, b = 2$), $h \equiv 0$, while one inter-boundary mode adjusts relative loop scale without interior cones.

\subsection{Spatial Design Controls}\label{sec:model-controls}

Because the two core Hodge factorizations $A^\top C A$ and $D^\top C^{-1}D$ depend solely on mesh geometry, three lightweight edits to Eq.~\eqref{eq:score} add localized controls on the same conformal model (Fig.~\ref{fig:controls}):
(1)~\emph{Hard regional cone constraints} fix $z_v \in \{0,\pm 1\}$ on marked vertices (e.g., $z_v = 0$ for cone-free regions), restricting the admissible charges.
(2)~\emph{Local direction weights and target offsets} replace the directional term with $\tfrac{w_D}{2}\min_{\phi_0}\|\gamma + \phi_0\mathbf{1} - \delta\|_{W_D}^2$ ($W_D = \mathrm{diag}(M_F w_F)$): because non-uniform $W_D$ or nonzero $\delta$ shifts the optimal global rotation away from zero, $\phi_0^*$ is eliminated analytically while keeping $D^\top C^{-1}D$ unchanged, and propagating $w_F \neq 1$ into the auxiliary smooth fit updates only the diagonal of $A_\theta$ on its existing sparsity pattern.
(3)~\emph{Local sizing/density targets} replace the scale term with $\tfrac{w_A}{2}\|u - u^*\|_{M_V}^2$; since surface quad density scales as $\rho \propto e^{2u}$, a target relative area density $\rho$ sets the $M_V$-centered target $u^* = \tfrac{1}{2}\log\rho$ without altering either Hodge factorization.

\section{Discrete Optimization with Reusable Responses}\label{sec:opt}

\begin{figure*}[t]
  \centering
  \includegraphics[width=0.93\textwidth]{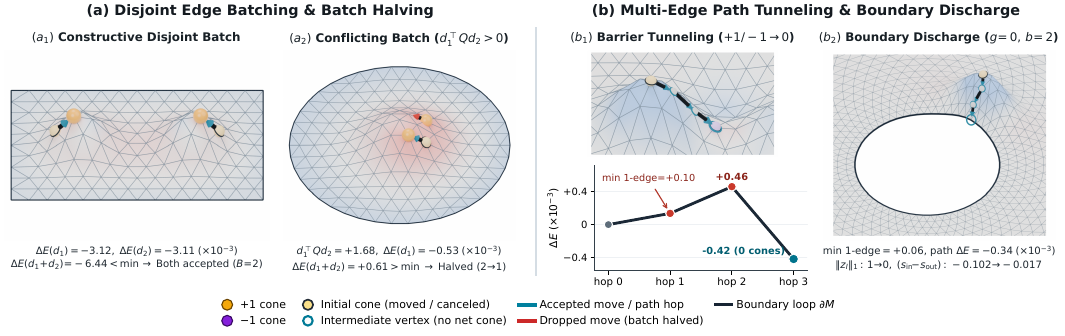}
  \caption{Discrete topological search moves on fixed-geometry prefactorizations. \textbf{(a)}~Disjoint edge batching with exact coupling verification via Eq.~\eqref{eq:quad_form}: separated moves reinforce and accept jointly~$(a_1)$, whereas positive cross-coupling ($d_1^\top Q d_2 > 0$) triggers dyadic batch halving~$(a_2)$. \textbf{(b)}~Nonlocal charge-transfer paths: a three-hop primal path tunnels across a single-edge barrier to annihilate an opposing $+1/-1$ cone pair~$(b_1)$, and an open-surface path ($g = 0, b = 2$) discharges an interior cone onto a boundary loop~$(b_2)$, adjusting relative loop scale.}
  \Description{Four-panel illustration of our discrete search moves: (a1) constructive disjoint edge batch accepting both moves, (a2) conflicting disjoint batch with positive cross-coupling triggering batch halving, (b1) three-hop barrier-tunneling path annihilating a +1/-1 cone pair with an energy-vs-hop plot, and (b2) boundary-discharge path clearing an interior cone onto a boundary loop on an annulus.}
  \label{fig:search_moves}
\end{figure*}

\subsection{Fixed-Branch Move Evaluation}\label{sec:opt-exact}

Because cone edits affect scale and alignment globally, we score moves with our response model rather than local heuristics. By Eq.~\eqref{eq:connection}, applying a finite integer edit $d \in \mathbb{Z}^{|E_I|}$ ($p \mapsto p + d$) shifts cross-edge discrepancies by $\Delta r = -qd$ and cone charges by $\Delta z_I = B_I d$. Since Eqs.~\eqref{eq:smooth} and \eqref{eq:open} (with $R{=}I$ and $A{=}B^\top$ when $b{=}0$) are linear in $r$, all four responses $(\Delta u, \Delta\gamma, \Delta h, \Delta\omega)$ depend linearly on $d$ on a fixed angle branch. The continuous part $E_{\mathrm{cont}}(p)$ of Eq.~\eqref{eq:score} is therefore quadratic in $p$ on a fixed branch. Writing $c(p) = \nabla_p E_{\mathrm{cont}}(p)$ for its gradient and $Q \succeq 0$ for the fixed-branch Hessian, any finite edit $d$ incurs the exact increment
\begin{equation}\label{eq:branch_inc}
\begin{aligned}
  \Delta E^{\mathrm{branch}}(d) = &\; c(p)^\top d + \frac{1}{2} d^\top Q d \\
    &+ w_C\bigl(\|z_I + B_I d\|_1 - \|z_I\|_1\bigr),
\end{aligned}
\end{equation}
\begin{equation}\label{eq:quad_form}
\begin{aligned}
  d^\top Q d = &\; w_A\|\Delta u\|_{M_V}^2 + w_D\|\Delta\gamma\|_{M_F}^2 \\
               &+ w_H\|\Delta h\|_{C^{-1}}^2 + \lambda_s\|\Delta\omega\|_W^2 \ge 0.
\end{aligned}
\end{equation}
A sizing target $u^*$ (Section~\ref{sec:model-controls}) shifts only $c(p)$, not $Q$.

Because $E_{\mathrm{cont}}(p)$ is a quadratic functional of three prefactored linear solves, a standard adjoint pass reuses those same three factorizations ($\le 3$ solves) to update the full edge gradient $c(p)$ at each accepted state without forming dense $Q$. Paired with single-edge self-costs $Q_{jj}$ (Section~\ref{sec:opt-schedule}), an $O(|E_I|)$ pass screens admissible signed single-edge moves $d = \sigma e_j$ ($\sigma \in \{-1,+1\}$) using their exact fixed-branch cost $\sigma c_j(p) + \tfrac{1}{2}Q_{jj} + w_C\Delta\|z_I\|_1$. Moves are accepted only after branch alignment confirms descent of Eq.~\eqref{eq:score}.

\subsection{Disjoint Edge Batches and Nonlocal Charge Paths}\label{sec:opt-moves}

A single-edge move $d = \sigma e_j$ shifts a cone across an edge or creates/cancels a dipole. Packing vertex-disjoint descending moves accelerates convergence, but long-range Poisson cross-couplings in $\tfrac{1}{2}(\sum_i d_i)^\top Q (\sum_i d_i) = \tfrac{1}{2}\sum_i d_i^\top Q d_i + \sum_{i<j} d_i^\top Q d_j$ can make moves conflict (Fig.~\ref{fig:search_moves}a). We greedily pack top vertex-disjoint moves into $d = \sum_i d_i$ and evaluate Eq.~\eqref{eq:quad_form} via one forward linear response (three solves), capturing all cross-couplings without forming $Q_{ij}$. If positive cross-coupling leaves the batch worse than the best single-edge move, batch halving iteratively halves the batch until it improves on or falls back to that move.

To tunnel across multi-edge energy barriers, we route directed primal shortest paths $d = \sum_{j \in \mathcal{P}} \sigma_j e_j$ along interior edges whose intermediate charges cancel. Because $B_I d$ is nonzero only at the endpoints, each path transfers unit charge between cones, low-potential interior vertices, or boundary loops (Fig.~\ref{fig:search_moves}b). This brings nonlocal charge transfer and dipole cancellation~\cite{Farchi2018IntegerOnly} into edge-matching space, where cones, handle holonomies, and target-relative alignment are scored jointly. On open meshes, terminating a path on a boundary loop performs \emph{boundary discharge}, clearing an interior cone while adjusting relative loop scales. Two equally oriented paths between the same endpoints transfer identical charge, but winding differently around a handle changes its holonomy according to their signed crossings with a dual generator, and the harmonic term in Eq.~\eqref{eq:score} scores the resulting mismatch. We guide Dijkstra path proposals with positive directed weights that combine Eq.~\eqref{eq:smooth}'s fixed-$\theta$ local increments and non-scale gradients, since conservative scale gradients telescope to the endpoints. Top paths are ranked by Eq.~\eqref{eq:branch_inc} and accepted only when they lower Eq.~\eqref{eq:score} after branch alignment. Single-edge moves, batches, and paths preserve all fixed regional charges from Section~\ref{sec:model-controls}; paths may still cross protected regions because their intermediate charges cancel.

\begin{table*}[t]
  \centering
  \caption{Curvature benchmark across 278 evaluated QuadWild meshes under \textsc{RSP}. Scale-only selectors are conformal-frame initialized; \textsc{IOQ} skips 99 meshes exceeding its 8\,GiB cap, and SparseIntCones has 9 upstream failures. Conv.\ counts converged \textsc{RSP} runs; Valid counts \textsc{RSP} maps passing all checks, and Poisson counts single-solve linear integrations passing the same checks (no \textsc{RSP} iteration). Medians summarize each method's valid \textsc{RSP} maps; paired $\Delta$ reports median Ours minus baseline on the $n$ jointly valid meshes (negative favors Ours).}
  \label{tab:main_results}
  \small
  \renewcommand{\arraystretch}{0.95}
  \setlength{\tabcolsep}{4.4pt}
  \begin{tabular}{@{}llcccrrrrrrr@{}}
    \toprule
    & & \multicolumn{2}{c}{\textsc{RSP}} & Poisson & \multicolumn{3}{c}{Medians (Valid)} & \multicolumn{4}{c}{Paired $\Delta$ (Ours $-$ Base.)} \\
    \cmidrule(lr){3-4}\cmidrule(lr){5-5}\cmidrule(lr){6-8}\cmidrule(lr){9-12}
    Method & Scope & Conv. & Valid & Valid & $A$ ($^\circ$) & $S$ & $C$ & $n$ & $\Delta A$ ($^\circ$) & $\Delta S$ & $\Delta C$ \\
    \midrule
    \textbf{Ours ($\beta = 0.5$)} & All 278 & 264 & \textbf{211} & \textbf{219} & 6.89 & 0.141 & 32 & --- & --- & --- & --- \\
    Input $G$ & All 278 & 241 & 195 & 150 & 4.22 & 0.197 & 38 & 192 & $+1.41$ & $-0.040$ & $-2$ \\
    Guided 4-RoSy seed & All 278 & 247 & 203 & 177 & 6.19 & 0.170 & 36 & 196 & $+0.15$ & $-0.020$ & $-2$ \\
    Kn\"oppel (unguided) & All 278 & 235 & 196 & 146 & 17.97 & 0.204 & 36 & 187 & $-10.07$ & $-0.054$ & $-4$ \\
    \midrule
    \textsc{IOQ} & Closed 254 (155 run) & 133 & 124 & 55 & 16.20 & 0.172 & 22.5 & 122 & $-8.54$ & $-0.035$ & $+2.5$ \\
    MoveIntCone ($d{=}0.2$) & Closed $g{=}0$ 118 & 112 & 89 & 91 & 17.75 & 0.167 & 12 & 83 & $-9.77$ & $-0.013$ & $+10$ \\
    SparseIntCones ($d{=}0.2$) & Closed $g{=}0$ 118 (109 run) & 105 & 86 & 83 & 17.51 & 0.151 & 13 & 83 & $-9.77$ & $-0.002$ & $+8$ \\
    \bottomrule
  \end{tabular}
\end{table*}

\subsection{Factorization Reuse, Selected Inversion, and Schedule}\label{sec:opt-schedule}

Move evaluation and self-costs rely on three one-time sparse $LDL^\top$ factorizations per mesh: $A^\top C A$, $D^\top C^{-1}D$, and $A_\theta$. Setting $d = e_j$ in Eq.~\eqref{eq:quad_form} expresses the scale and angle self-costs $Q_{jj}$ as squared-inverse quadratic forms $a^\top \mathcal{L}^{-1} M_r \mathcal{L}^{-1} a$ (up to one mean-centering solve), where $\mathcal{L}$ is the pinned SPD submatrix of $A^\top C A$ or $D^\top C^{-1} D$ and $M_r$ is its restricted diagonal mass matrix (from $R^\top M_V R$ or $M_F$). Standard selected inversion (\textsc{SelInv}, \citealp{Lin2011SelInv}) exposes sparse entries of $\mathcal{L}^{-1}$, not $\mathcal{L}^{-1} M_r \mathcal{L}^{-1}$. Observing that
\begin{equation}\label{eq:mass_deriv}
  a^\top \mathcal{L}^{-1} M_r \mathcal{L}^{-1} a
  = -\left.\frac{\partial}{\partial \tau}\, a^\top (\mathcal{L} + \tau M_r)^{-1} a \right|_{\tau = 0},
\end{equation}
and that diagonal $M_r$ preserves the sparsity pattern of $\mathcal{L}$, we precompute all $|E_I|$ self-costs $Q_{jj}$ on the existing Cholesky pattern from selected inverse entries and their mass derivatives (differentiated \textsc{SelInv}, \citealp{Andersen2013Barriers}). For $|E_I| \ge 6{,}000$, we cache $Q_{jj}$ this way; uncached entries use direct solves.

Search starts from two complementary seeds: direct rounding of $G$ and a guided 4-RoSy seed, which smooths the fourfold complex representation $e^{4i\alpha}$ of $G$ using parallel transport and rounds the resulting field to an integer matching. Both seeds are repaired to unit charges $z_I \in \{-1,0,1\}^{|V_I|}$ via greedy adjacent-edge edits and spanning-tree flow while honoring fixed regional charges. Under a shared $B_{\max} = 512$ operation cap, each seed runs up to $192$ operations before the lower-energy seed continues unless stalled (Algorithm~\ref{alg:search}).

\begin{algorithm}[t]
\caption{Direction-guided discrete topology optimization.}\label{alg:search}
\footnotesize
\begin{algorithmic}[1]
\Require Mesh $\mathcal{M}$, target $G$ ($\alpha$), preference $\beta$, cap $B_{\max}{=}512$ (opt.\ controls).
\Ensure Branch-aligned $(p, z_I, u, \alpha{+}\theta)$ with $z_I \in \{-1,0,1\}^{|V_I|}$.
\State Factor $A^\top C A$, $D^\top C^{-1}D$, $A_\theta$ once ($LDL^\top$); cache $Q_{jj}$ via differentiated \textsc{SelInv} using Eq.~\eqref{eq:mass_deriv} for $|E_I| \ge 6{,}000$ (direct solves otherwise).
\State Build direct-$G$ and guided 4-RoSy seeds; repair to unit charges, branch-align via Eq.~\eqref{eq:smooth}, and score against $G$ via Eq.~\eqref{eq:score}.
\For{each seed ($\le 192$ operations), then lower-energy seed if not stalled (total $\le B_{\max}$)}
  \State Update continuous gradient $c(p)$ via one adjoint pass ($\le 3$ solves).
  \State Screen single-edge moves via $Q_{jj}$; verify disjoint batch via Eq.~\eqref{eq:quad_form} with halving.
  \State \textbf{if} branch-aligned batch lowers Eq.~\eqref{eq:score} \textbf{then} accept and \textbf{continue}.
  \State Propose primal paths (incl.\ boundary discharge); rank by Eq.~\eqref{eq:branch_inc}.
  \State \textbf{if} top branch-aligned path lowers Eq.~\eqref{eq:score} \textbf{then} accept; \textbf{else} mark stalled, \textbf{break}.
\EndFor
\State \Return Lowest-energy branch-aligned $(p, z_I, u, \alpha{+}\theta)$.
\end{algorithmic}
\end{algorithm}

\section{Evaluation}
\label{sec:eval}

\subsection{Benchmarks, Parameterization Backend, and Validity}
\label{sec:eval-setup}

We evaluate on 278 single-component orientable QuadWild manifolds~\cite{Pietroni2021Reliable} spanning $480$--$100\text{k}$ faces ($118$ closed $g = 0$, $136$ closed $1 \le g \le 133$, $24$ open $1 \le b \le 15$) and $50$ raw Thingi10K meshes~\cite{Zhou2016Thingi10K} with aspect ratio $\ge 10$ and $2\text{k}$--$50\text{k}$ faces. Default parameters are set on QuadWild-24, a fixed 24-mesh subset (8 closed $g = 0$, 10 closed $g \ge 1$, 6 open) also used for ablations. For the curvature benchmark, we compute $G$ from per-face principal curvature directions via a connection Laplacian; Input~$G$ rounds parallel-transported cross-edge angle differences to the nearest quarter-turn. Unless noted otherwise, all runs share one default parameter set and the official fixed-topology \textsc{RSP} backend~\cite{Corman2025RSP} against original $G$; Section~\ref{sec:eval-baselines} also evaluates an implementation of \textsc{RSP}'s progressive cut-and-glue strategy atop those routines.

\begin{figure*}[t]
  \centering
  \includegraphics[width=0.93\textwidth]{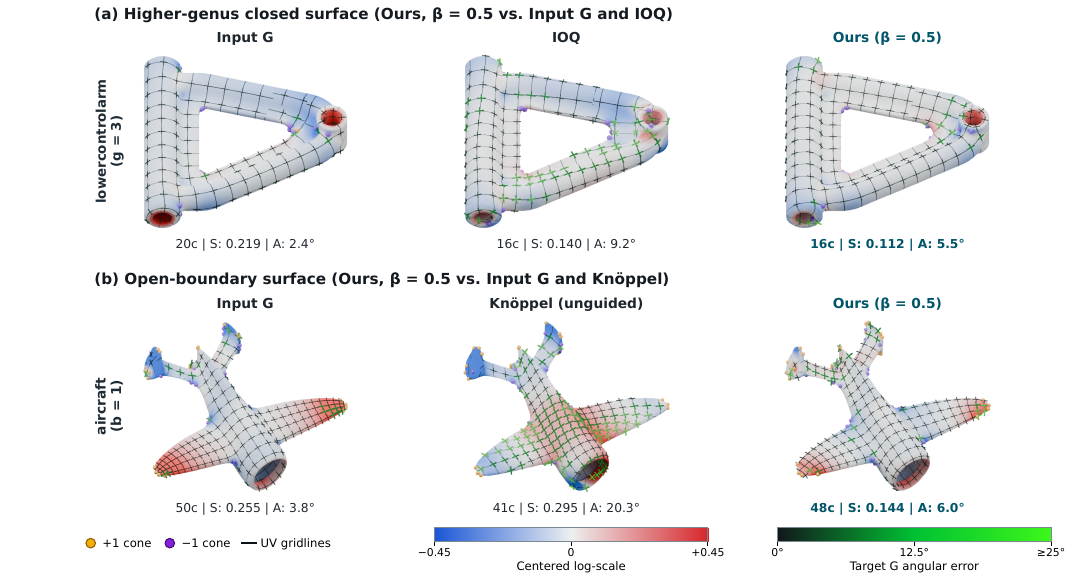}
  \caption{Cross-topology comparisons under \textsc{RSP}. \textbf{(a)}~On a genus-$3$ surface, our default lowers both $S$ and $A$ relative to \textsc{IOQ} at equal $16$ cones and cuts $S$ relative to Input~$G$ at higher $A$. \textbf{(b)}~On an open shell, it shows the same trade-off against Input~$G$ and improves both metrics over Kn\"oppel (unguided).}
  \Description{Two rows show a genus-3 suspension arm and an open aircraft shell. Each row places two comparison methods on the left and our result on the right. Blue-to-red surface shading shows centered log-scale, and dark-to-green crosses show increasing angular error.}
  \label{fig:comparison}
\end{figure*}

We compare publicly available orthogonal cross-field topology selectors on their supported subsets: scale-only Move\-Int\-Cone~\cite{Du2026EfficientIntegerCones} and Sparse\-Int\-Cones~\cite{Zhang2024PracticalIntegerCones} on closed $g = 0$ (9 upstream failures for Sparse\-Int\-Cones), and \textsc{IOQ}~\cite{Farchi2018IntegerOnly} on closed meshes (skipping $99/254$ exceeding its 8\,GiB memory cap). On all 278 meshes, we compare Input~$G$, Kn\"oppel (unguided, \citealp{Knoppel2013GloballyOptimal}), our repaired guided 4-RoSy seed (scale-agnostic complex smoothing), and our method; \textsc{IOQ} and the guided seed share our Eq.~\eqref{eq:smooth} starting frame $\alpha + \theta$. Scale-only selectors output only charges $z$, so we initialize them in \textsc{RSP} with their integrable conformal frame (globally aligned to $G$) to reduce $90^\circ$ wrapping. An \textsc{RSP} map is \emph{valid} iff \textsc{RSP} converges with zero flips or collapses, UV cone windings matching $z_I$, and verified integer loop holonomies. We report area-weighted two-axis RMS angular error $A$ ($\bmod\,\pi/2$), centered log-scale RMS $S$, and interior cone count $C$, plus \emph{Poisson integration}: single-solve linear integration~\cite{Kalberer2007QuadCover,Bommes2009MIQ} of each coframe scaled by a common linear conformal scale reconstructed from its topology, checked under the same validity criteria.

\subsection{External Comparisons and Matched-Budget Analysis}
\label{sec:eval-baselines}

On the 278 QuadWild meshes (Table~\ref{tab:main_results}), our default produces 211 valid \textsc{RSP} maps versus 195--203 for the cross-field baselines, and 219 valid Poisson maps versus 146--177. Because \textsc{RSP} adds anisotropic stretch without a flip barrier, all 53 converged but invalid \textsc{RSP} maps for our method contain local flips (zero collapses, median three flipped faces), so single-solve Poisson succeeds on some meshes where \textsc{RSP} flips. On the 192 meshes valid for both our method and Input~$G$, paired medians trade $+1.41^\circ$ angular error for two fewer cones and $0.040$ lower log-scale distortion (Fig.~\ref{fig:comparison}b), consistent with the 254 unseen meshes outside QuadWild-24 (191 vs.\ 176 valid). Against the guided 4-RoSy seed, our default adds eight valid maps, reduces median $S$ by $0.020$, and removes two cones at $+0.15^\circ$ angular cost. It also improves all three metrics over Kn\"oppel (unguided) and lowers both $A$ and $S$ over \textsc{IOQ} overall and at equal 16 cones (Fig.~\ref{fig:comparison}a). On QuadWild-24, raw-$G$ starts leave \textsc{RSP} validity and distortion unchanged for both Input~$G$ and our method while raising our median Newton iterations from 7 to 12. Gains thus come from topology rather than warm-start smoothing.

\begin{table}[t]
  \centering
  \caption{Closed genus-$0$ matched-budget comparison on $118$ QuadWild meshes (Ours $\beta = 0$ vs.\ conformal-initialized MoveIntCone and SparseIntCones, $90\%$--$100\%$ cone band; Ours $-$ ref.). Rows~1--4 average across selectors per mesh; the bottom row pools exact equal-cone curvature pairs across $81$ meshes.}
  \label{tab:matched_budget}
  \footnotesize
  \renewcommand{\arraystretch}{0.98}
  \setlength{\tabcolsep}{0pt}
  \begin{tabular*}{\columnwidth}{@{\extracolsep{\fill}}lrrr@{}}
    \toprule
    Target \& budget & Lower $A$ & Med.\ $\Delta A$ & Med.\ $\Delta S$ \\
    \midrule
    Curvature, $d = 0.1$ & $58/68$ meshes & $-1.88^\circ$ & $+0.0001$ \\
    \textbf{Curvature, $d = 0.2$} & $\mathbf{64/77}$ \textbf{meshes} & $\mathbf{-1.43^\circ}$ & $\mathbf{-0.0009}$ \\
    Curvature, $d = 0.4$ & $39/61$ meshes & $-0.66^\circ$ & $-0.0019$ \\
    Sinusoidal twist, $d = 0.2$ & $64/78$ meshes & $-1.15^\circ$ & $0.0000$ \\
    \midrule
    Exact equal cones (all $d$) & $213/293$ pairs & $-0.83^\circ$ & $-0.0004$ \\
    \bottomrule
  \end{tabular*}
\end{table}

At their default distortion target $d = 0.2$, the scale-only selectors Move\-Int\-Cone and Sparse\-Int\-Cones use fewer cones than our default in Table~\ref{tab:main_results}. To compare at similar counts on the 118 closed genus-$0$ meshes, we sweep eight cone penalties at $\beta = 0$ and, before running \textsc{RSP}, pair each reference result with our highest-count candidate within $90\%$--$100\%$ of its cone count, averaging paired differences per mesh across valid pairs. Whereas Fig.~\ref{fig:teaser}a illustrates our balanced $\beta = 0.5$ operating point at an equal $20$-cone count, Table~\ref{tab:matched_budget} quantifies scale-priority $\beta = 0$ at matched budgets and near-zero median scale change across $d \in \{0.1, 0.2, 0.4\}$, the sinusoidally twisted field, and exact equal cones; gains narrow at $d = 0.4$, where most pairs hit the eight-cone floor.

\begin{figure*}[t]
  \centering
  \includegraphics[width=0.83\textwidth]{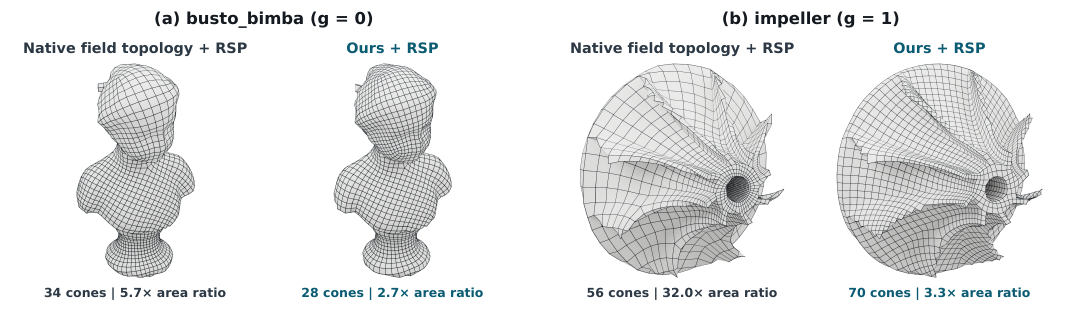}
  \caption{Flip-free pure-quad meshes extracted from \textsc{RSP}'s built-in curvature field via quantization~\cite{CoudertOsmont2024Quantization} and \texttt{libQEx}~\cite{Ebke2013QEx} at matched target quad counts. Optimizing cones at $\beta = 0.5$ before \textsc{RSP} lowers the quad-area ratio $p_{95}/p_{05}$ on both examples, cutting cones $34 \to 28$ in \textbf{(a)} and removing all eight valence-2 vertices ($56 \to 70$ cones) in \textbf{(b)}.}
  \Description{Two pairs of pure-quad meshes show a bust on the left and an impeller on the right. Each pair places Native field topology + RSP on the left and our cone-optimized RSP result on the right, with labels reporting cone counts and quad-area ratios.}
  \label{fig:quads}
\end{figure*}

\begin{figure*}[t]
  \centering
  \includegraphics[width=0.83\textwidth]{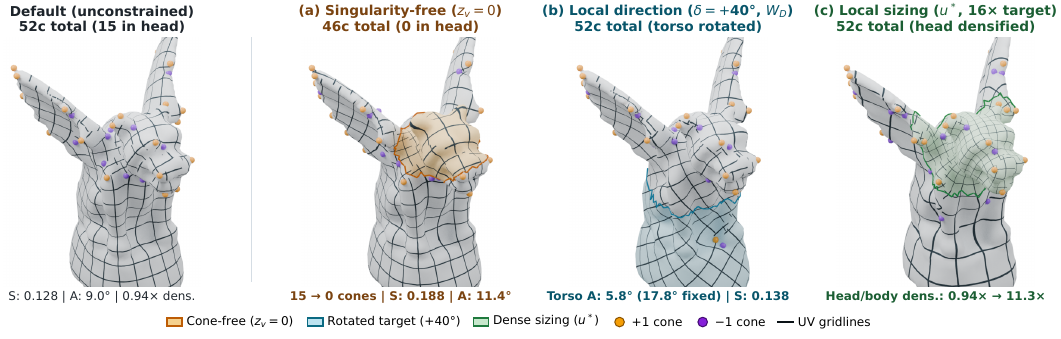}
  \caption{Spatial controls on a gargoyle relative to the unconstrained default (left). \textbf{(a)}~Head cone exclusion clears all 15 head singularities. \textbf{(b)}~After a $+40^\circ$ torso edit, reoptimizing at $52$ cones gives $5.8^\circ$ torso error; frozen cones leave $17.8^\circ$ from the original frame, or $14.9^\circ$ after refitting the start. \textbf{(c)}~At equal $52$ cones, a $16\times$ head target ($\beta = 0.25$) raises head/body UV density $0.94\times \to 11.3\times$ using only the cone selector.}
  \Description{Four views of a winged gargoyle show the default cone placement, a head cleared of cone markers, a rotated target on the torso, and a denser UV grid on the head.}
  \label{fig:controls}
\end{figure*}

\paragraph{Progressive cut-and-glue \textsc{RSP}}
We also implemented \textsc{RSP}'s progressive cut-and-glue strategy~\cite[Sec.~5.3.2]{Corman2025RSP} atop the public fixed-topology routines on the 12 QuadWild-24 curvature meshes with $|F| \le 22\text{k}$. Re-evaluating topologies in the shared fixed-topology backend at comparable cone counts yields similar quality (Ours $9/12$ vs.\ $7/12$ valid), whereas $40$--$170$ repeated full-mesh KKT--Newton solves make cut-and-glue $4\times$--$52\times$ slower single-threaded and time out at $1{,}800\,\mathrm{s}$ on denser shapes like Fig.~\ref{fig:teaser}b (vs.\ $16.4\,\mathrm{s}$ for ours).

\paragraph{Downstream quad extraction.}
On 12 closed QuadWild surfaces spanning genera $0$--$9$, we compare \textsc{RSP}'s native rounded curvature topology against optimizing cones at $\beta = 0.5$ before \textsc{RSP} under identical quantization~\cite{CoudertOsmont2024Quantization} and \texttt{libQEx}~\cite{Ebke2013QEx} at matched target quad counts. Both yield flip-free pure-quad meshes on 11/12 surfaces; across all 10 shared successes, our cones cut the quad-area ratio $p_{95}/p_{05}$ (median $5.1\times \to 2.8\times$) and mean aspect ratio on every mesh with comparable mean scaled Jacobian ($0.988$ vs.\ $0.989$), at higher UV angular error. In Fig.~\ref{fig:quads}, this relieves head-to-pedestal disparity on the bust~(a) with fewer cones, and hub-to-rim disparity on the impeller~(b), removing all eight valence-2 vertices and lifting minimum scaled Jacobian from $0.035$ to $0.360$.

\subsection{Model Validation, Design Fields, and Spatial Controls}
\label{sec:eval-controls}

To compare the two directional scores, we sweep $w_D$ independently for $\gamma$ and $\theta$ from the same initial matchings on six meshes, each with a curvature field and a smooth sinusoidal twist. Because $\theta$ damps low-frequency swirl differently across meshes, equalizing the initial directional terms requires $1.5\times$--$89\times$ weight shifts across meshes and seeds, making a single weight inconsistent for $\theta$. Under the common default weight, $\gamma$ lowers final UV angular error on $8$ of $12$ mesh--target pairs (median $-0.37^\circ$); after per-condition weight tuning, the two scores reach the same median error ($0.00^\circ$) at equal cone counts and similar log-scale distortion. Scoring $\gamma$ thus delivers comparable tuned quality without per-mesh weight calibration.

Decoupling $G$ from curvature on QuadWild-24 amplifies our gains on both a sinusoidal twist and a projected-axis field. On the projected-axis field, our default produces 20/24 valid \textsc{RSP} maps versus 18/24 for the guided seed and 16/24 for Input~$G$. On 15 pairs with Input~$G$, it cuts median $A$ by $2.02^\circ$, $S$ by $0.179$, and $C$ by $53$; against the guided seed (18 pairs), it matches alignment ($-0.01^\circ$) at $0.026$ lower $S$ and six fewer cones. On the sinusoidal twist, direct rounding preserves the original signed cones on 21/24 meshes, while our default relocates cones on 19 of those 21 and reduces $A$ by $0.39^\circ$ and $S$ by $0.062$ across 18 pairs with Input~$G$.

\begin{figure*}[t]
  \centering
  \includegraphics[width=0.89\textwidth]{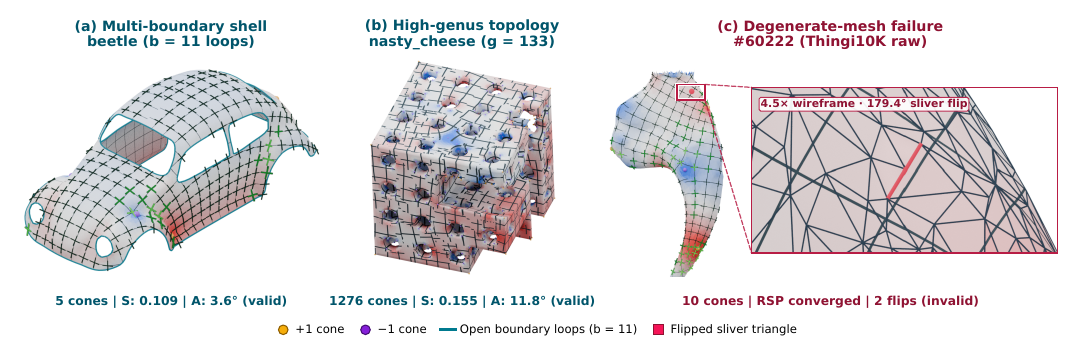}
  \caption{Extreme topologies and a raw-mesh failure under \textsc{RSP}. Without special casing, our method produces valid maps on \textbf{(a)}~an $11$-boundary car shell (loops in teal) and \textbf{(b)}~a genus-$133$ closed surface. \textbf{(c)}~On a raw Thingi10K mesh, \textsc{RSP} converges globally yet leaves two flipped triangles; the inset highlights a nearly $180^\circ$ inverted sliver (color scales follow Fig.~\ref{fig:comparison}).}
  \Description{Single-row comparison showing an 11-boundary car shell, a genus-133 perforated block, and a raw Thingi10K mesh paired side-by-side with a 4.5x wireframe close-up highlighting a 179.4-degree inverted sliver in red.}
  \label{fig:generality}
\end{figure*}

\begin{table}[t]
  \centering
  \caption{Model-term and $\beta$ ablation on QuadWild-24 under \textsc{RSP}. Medians use valid maps; paired $\Delta$ is vs.\ $\beta{=}0.50$ (median paired $\Delta C{=}0$ in all rows). $w_D{=}0$ holds $w_A{=}10^{-0.5}$; for $g{>}0$, $\beta{=}0.50$ yields $9/10$ valid.}
  \label{tab:model_ablation}
  \footnotesize
  \renewcommand{\arraystretch}{0.95}
  \setlength{\tabcolsep}{0pt}
  \begin{tabular*}{\columnwidth}{@{\extracolsep{\fill}}lcrrrr@{}}
    \toprule
    & & \multicolumn{2}{c}{Med.\ (Valid)} & \multicolumn{2}{c}{Paired $\Delta$} \\
    \cmidrule(lr){3-4}\cmidrule(lr){5-6}
    Setting & Valid & $S$ & $A$ ($^\circ$) & $\Delta S$ & $\Delta A$ ($^\circ$) \\
    \midrule
    $\beta = 0.00$ ($100{:}1$) & $19/24$ & $0.103$ & $10.1$ & $-0.031$ & $+2.85$ \\
    $\beta = 0.25$ ($10{:}1$)  & $19/24$ & $0.112$ & $8.6$  & $-0.024$ & $+1.90$ \\
    $\boldsymbol{\beta = 0.50}$ \textbf{(def.)} & $\mathbf{20/24}$ & $\mathbf{0.136}$ & $\mathbf{6.5}$ & --- & --- \\
    $\beta = 0.75$ ($1{:}10$)  & $19/24$ & $0.159$ & $4.8$  & $+0.026$ & $-1.33$ \\
    $\beta = 1.00$ ($1{:}100$) & $21/24$ & $0.178$ & $4.7$  & $+0.037$ & $-2.06$ \\
    \midrule
    $w_D = 0$ (dir.\ off)      & $17/24$ & $0.121$ & $6.2$  & $-0.007$ & $+1.26$ \\
    $w_H = 0$ ($g > 0$)        & $7/10$  & $0.137$ & $8.1$  & $-0.005$ & $+0.24$ \\
    $\lambda_s = 0$ (smooth.\ off) & $17/24$ & $0.126$ & $6.9$ & $-0.006$ & $-0.40$ \\
    \bottomrule
  \end{tabular*}
\end{table}

Sweeping $\beta$ from $0$ to $1$ trades angular alignment against scale uniformity on Fig.~\ref{fig:teaser}b and QuadWild-24 (Table~\ref{tab:model_ablation}), with steady validity and zero median cone change. Disabling the direction term ($w_D = 0$) lowers validity and increases paired angular error. Disabling the smoothness regularizer ($\lambda_s = 0$) lowers $S$ by $0.006$ and $A$ by $0.40^\circ$ on valid meshes, but drops validity from $20/24$ to $17/24$, nearly doubles total flipped triangles ($13 \to 25$), and breaks the high-genus shapes ($g = 51, 133$). Across 11 stratified meshes (up to 18 candidates near the default cone count), our physical score tracks \textsc{RSP} distortion $\tfrac{1}{2}w_A S^2 + \tfrac{1}{2}w_D A_{\mathrm{rad}}^2$ with median Spearman $\rho = 0.977$ ($0.973$ equal-cone). With default $\lambda_s = 10^{-3}$, the full score retains $\rho = 0.882$ ($0.917$ equal-cone) and selects a top-two candidate on 10/11 meshes; outliers arise under strong anisotropy or multi-boundary scale variation. Fig.~\ref{fig:controls} demonstrates regional cone exclusion, torso rotation, and head densification using the same Hodge factors.

\subsection{Topological Generality, Runtime Scaling, and Raw-Mesh Failures}
\label{sec:eval-generality}

\begin{figure}[b]
  \centering
  \includegraphics[width=0.80\columnwidth]{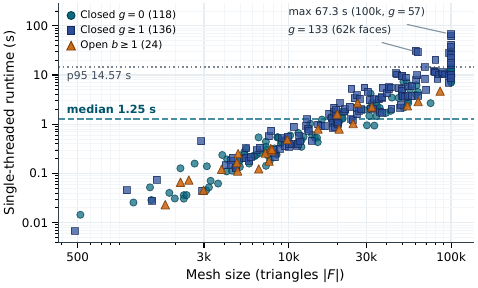}
  \caption{Single-threaded runtime across all 278 QuadWild meshes (median $1.25\,\mathrm{s}$; 95th percentile $14.57\,\mathrm{s}$; max $67.3\,\mathrm{s}$).}
  \Description{Log-log scatter plot of single-threaded solver runtime versus triangle count across all 278 evaluated QuadWild meshes grouped by closed genus-zero, higher-genus closed, and open-boundary topologies, with dashed reference lines marking the 1.25-second median and 14.57-second 95th percentile.}
  \label{fig:runtime}
\end{figure}

\paragraph{Extreme topologies.}
Without special casing, Eq.~\eqref{eq:open} handles an $11$-boundary shell and a genus-$133$ surface (Fig.~\ref{fig:generality}a--b); disabling $w_H$ (Table~\ref{tab:model_ablation}) drops positive-genus validity on QuadWild-24 from $9/10$ to $7/10$, breaking the genus-$133$ case.

\paragraph{Runtime and search ablations.}
On QuadWild, setup takes $0.37\,\mathrm{s}$ median for the three sparse $LDL^\top$ factorizations and self-cost caching, and single-threaded Eigen C++ runtime is $1.25\,\mathrm{s}$ median (Fig.~\ref{fig:runtime}). In QuadWild-24 search ablations, differentiated \textsc{SelInv} provides a $1.99\times$ median speedup with identical matchings on all $24$ meshes. Omitting local edge moves raises median objective by $+3.91\%$ and slows search $3.96\times$, while removing disjoint batching multiplies steps by $2.46\times$ and stalls on large high-genus meshes. Omitting nonlocal paths or either initial seed degrades $8$--$13$ meshes.

\paragraph{Failures on raw Thingi10K meshes.}
On the $50$ raw Thingi10K stress meshes, our cone selection completes on every mesh and slightly improves \textsc{RSP} outcomes over cross-field baselines: $13$ converged runs and $3$ valid maps vs.\ $5$--$11$ and $1$--$2$. Validity remains low because severe triangle degeneracy limits parameterization: Fig.~\ref{fig:generality}c shows a nearly $180^\circ$ sliver flipping despite \textsc{RSP} convergence.

\section{Discussion and Limitations}
\label{sec:discussion}

Separating cross-edge discrepancy into conformal scaling, unabsorbable swirl, and handle mismatch lets us score candidate topologies with a linear conformal model before UV optimization. Because the Hodge operators depend only on the mesh geometry, discrete search and spatial controls both reuse the same factorizations without nested nonlinear solves.

\paragraph{Limitations.}
First, our linear score assumes isotropic scaling, so anisotropic stretch in \textsc{RSP} can alter candidate rankings. On open surfaces, free boundary directions with one scale per loop leave boundary or crease alignment unconstrained, and varying loop scale requires weight-dependent refactorization. Second, Eq.~\eqref{eq:branch_inc} is exact only on a fixed angle branch, and greedy descent reaches local rather than global optima. Results thus depend on initial matchings and take more steps on fine or high-genus meshes. Third, optimizing a pre-integration proxy cannot guarantee flip-free parameterization or quantization, leaving residual inversions on both clean and raw meshes with nearly $180^\circ$ slivers.

\paragraph{Future Work.}
Natural extensions include anisotropic response models, multilevel or global discrete solvers, explicit feature-crease or boundary alignment, and pairing our free-boundary model with free-boundary parameterizers. For raw meshes, pairing our selector with intrinsic triangulations~\cite{Sharp2019Navigating} is promising, as intrinsic edge flips enable robust discrete conformal maps~\cite{Gillespie2021DiscreteConformal}.

\clearpage
\bibliographystyle{ACM-Reference-Format}
\bibliography{references}
\end{document}